# Generalized many-body approach for near-field radiative heat transfer between nonspherical dipoles

Lindsay P. Walter and Mathieu Francoeur*

[1]Department of Mechanical Engineering, The University of Utah, Salt Lake City, UT 84112, USA

## ABSTRACT

A generalized fluctuational electrodynamics-based many-body approach for calculating near-field radiative heat transfer (NFRHT) between nonspherical dipoles is proposed. The geometric parameters of nonspherical dipoles are implemented in the definition of the self-term of the free-space Green's function. Dipole polarizability is defined a posteriori from the free-space Green's function solution such that polarizability calculation is an optional post-processing step rather than a required input. Both strong and weak forms of the generalized many-body approach are presented. It is shown that the approximate weak form is less computationally expensive but is only applicable to small particles characterized by size parameters less than ~0.24. The generalized many-body method is compared against an analytical solution for NFRHT between two spheroidal dipoles. A good agreement is obtained, and the small discrepancies are ascribed to differences in approximations for multiple reflections. The generalized many-body method is then applied to analyze the near-field spectral conductance between two SiC ellipsoidal dipoles. Results reveal that changes in the orientation of one of the ellipsoidal dipoles lead to active tuning of localized surface phonon resonance by up to three orders of magnitude. Finally, the spectral radiative thermal conductivity of a metamaterial composed of 1000 $SiO_2$ ellipsoidal particles is studied. The metamaterial displays anisotropic radiative thermal conductivity, with resonance values differing by up to a factor of 2.8 between different directions. The generalized many-body model of NFRHT presented in this paper may be used to develop particle-based metamaterials with novel, engineered radiative thermal properties.

---

*Corresponding author: mfrancoeur@mech.utah.edu

## I. INTRODUCTION

Localized surface phonons (LSPhs) are modes that arise from the coupling of electromagnetic waves with mechanical vibrations within a polar dielectric material of confined spatial dimension smaller than the characteristic wavelength. The particle analog of surface phonon-polaritons, LSPhs are dependent on the material and geometry of the confining domain and may be thermally generated in many polar dielectrics at room temperature. Like surface phonon-polaritons, the electromagnetic field associated with LSPh modes decays exponentially from the surface of the particle. For LSPhs to interact and couple between thermal objects, the objects must be separated by distances less than the thermal photon wavelength. This range of subwavelength separation distances is known as the near-field regime of thermal radiation.

Near-field radiative heat transfer (NFRHT), characterized by nontrivial electromagnetic wave dynamics and heat transfer exceeding Planck's blackbody limit [1], is modeled via fluctuational electrodynamics [2]. One NFRHT model for systems containing large numbers of particles is the many-body approach for dipoles [3]. In the dipole limit, particles must be small compared to the thermal photon wavelength and must be separated by distances greater than a few times the particle size [4]. While there exist other models of NFRHT between nonspherical particles that are not constrained to dipole approximations (e.g., Refs. [5–10]), these models are usually much more computationally demanding than dipole models. For example, one of the most computationally efficient methods for modeling NFRHT between arbitrarily shaped objects is the fluctuating-surface-current method [5,6]. Even with the reduced computational costs associated with the fluctuating-surface-current method, however, modeling large systems containing upwards of hundreds of arbitrarily shaped particles is impractical, and often intractable. For such large systems, dipole approximations are a better choice. Additionally, many-body approaches of NFRHT based on a general scattering formalism for thermal objects of arbitrary size and separation distance embedded in vacuum [11,12] and in a nonabsorbing background medium [13] have been developed and applied to systems of up to three objects. The scattering framework was recently extended to model heat transfer via electromagnetic waves in dissipative media and applied to predict the radiative conductivity of doped Si and $SiO_2$ [14]. In the remainder of this manuscript, the discussion will focus on the many-body approach for NFRHT between dipoles.

Since its development, the many-body approach in the dipole limit has been used to model NFRHT between small groups of less than five particles [15,16], to chains of more than ten particles [17,18], and to large two- and three-dimensional particle arrays with hundreds of particles [17-23]. In all of these works, however, the systems have been restricted to spherical particles. Other researchers have defined NFRHT models for nonspherical particles in the dipole limit, but these models have not yet achieved the same system scale as the many-body approach with hundreds of spherical particles. For example, researchers have developed NFRHT models for two and three spheroidal particles in the dipole limit [24-27] and models for thermally generated surface phonon-polaritons propagating along chains of spheroidal particles due to LSPh coupling [28]. The work presented in Ref. [28], however, is not a fluctuational electrodynamics model and therefore lacks the full heat transfer capabilities of many-body approaches. There is only one known work that describes a model of NFRHT for dipoles of greater geometric asymmetry than spheroids: the ellipsoidal dipole model developed by Nikbakht [25]. While defined in theory for any number of ellipsoidal dipoles, the system studied by Nikhakht is limited to a solitary ellipsoidal dipole interacting with the environment. Additionally, particle self-interaction is defined using approximative correction factors that obscure the role of geometric effects on NFRHT. As such, there is a need for nonspherical dipole models of NFRHT that are exact in their formulation, that are generalizable to a variety of dipole geometries, and that are of greater scale on par with that of the existing many-body approach for spherical dipoles.

In this paper, we present a generalized many-body approach to calculate NFRHT for large systems of nonspherical dipoles. Similar to the original many-body approach for spherical dipoles [3], this method is developed from the volume integral solution of Maxwell's equations and results in a self-consistent system Green's function equation. Instead of defining dipole polarizabilities *a priori* as done in fluctuational electrodynamics-based many-body approaches, the method developed in this paper maintains a Green's function description throughout. As such, this new method is similar to the many-body method for spherical dipoles embedded in an inhomogeneous environment based on the quantum Langevin framework [29]. The approach in this paper is also analogous to the discrete system Green's function method for modeling NFRHT between discretized objects of arbitrary shape [9] and results in a mathematically clear depiction of how geometric factors come into play for nonspherical dipoles. In this way, the generalized many-body approach presented here may be used to efficiently model NFRHT in large-scale particle devices, such as particle-based metamaterials with novel thermal transport behavior.

The rest of this paper is organized as follows. First, the generalized many-body approach is derived in Sec. II, where strong and weak forms of the method are distinguished. Equations for dipole polarizability that can be integrated into existing many-body approaches are derived *a posteriori* in Sec. III, and polarizabilities are compared as a function of particle size. Next, the generalized many-body approach is verified against an analytical solution for two spheroidal dipoles (Sec. IV.A). Finally, the generalized many-body approach is applied to study the thermal conductance between two SiC ellipsoidal particles of variable configuration (Sec. IV.B), to assess the accuracy of the weak form when calculating NFRHT between two SiC ellipsoidal particles (Sec. IV.C), and to analyze the radiative thermal conductivity in a metamaterial constructed of $SiO_2$ ellipsoidal particles (Sec. IV.D). Conclusions are provided in Sec. V.

## II. DESCRIPTION OF THE GENERALIZED MANY-BODY FORMALISM

The generalized many-body approach is based on fluctuational electrodynamics and is applicable to a system of $N$ thermal particles in the dipole limit embedded in a nonabsorbing background reference medium. The dipoles may be nonspherical and of variable geometry across the system (see Fig. 1). In the dipole limit, the characteristic length $L_{\text{ch}}$ of each particle must be much smaller than the thermal photon wavelength $\lambda_T$ defined by Wien's law, which is ~10 μm at room temperature (i.e., $L_{\text{ch}} \ll \lambda_T = 2898/T$ μm·K). Additionally, the center-of-mass separation distance $d$ between particles must be sufficiently large compared to the characteristic length of all particles (i.e., $d \gtrsim 3L_{\text{ch}}$) [4].

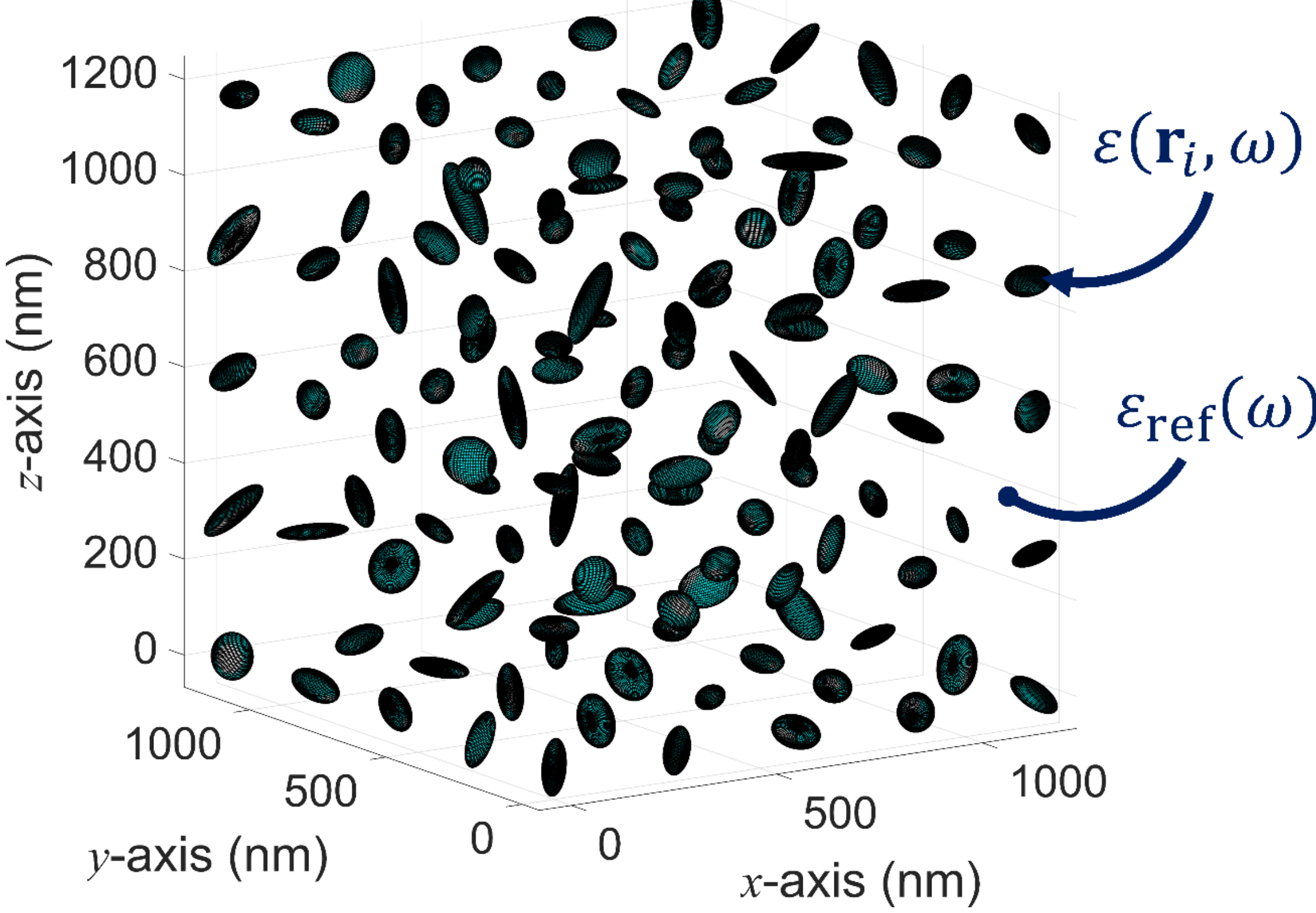

FIG. 1. System of many nonspherical dipoles of dielectric function $\varepsilon(\mathbf{r}_i, \omega)$ embedded in a nonabsorbing background reference medium described by a real-valued dielectric function $\varepsilon_{\mathrm{ref}}(\omega)$.

### A. Near-field radiative heat transfer

The total power dissipated in the $i$th particle due to thermal radiation exchanged with a group of $N$ particles may be defined as [9]

$$Q_{i,\mathrm{total}} = \frac{1}{2\pi}\int_0^\infty Q_i(\omega)\, d\omega = \frac{1}{2\pi}\int_0^\infty \sum_{\substack{j=1 \\ j\neq i}}^{N} \left[\Theta(\omega, T_j) - \Theta(\omega, T_i)\right] \mathcal{T}_{ij}(\omega)\, d\omega, \tag{1}$$

where $Q_i(\omega)$ is the spectral power dissipated in particle $i$ at angular frequency $\omega$, $\Theta(\omega, T_i) = \hbar\omega\left(\mathrm{e}^{\frac{\hbar\omega}{k_B T_i}} - 1\right)^{-1}$ is the mean energy of an electromagnetic state for a dipole at temperature $T_i$, $\hbar$ is the reduced Planck constant, $k_B$ is the Boltzmann constant, and $\mathcal{T}_{ij}(\omega)$ is the spectral transmission coefficient between the $i$th and $j$th dipoles. The spectral transmission coefficient may be written as

$$\mathcal{T}_{ij}(\omega) = 4k_0^4 \Delta V_i \Delta V_j \mathrm{Im}[\varepsilon(\mathbf{r}_i, \omega)] \mathrm{Im}\left[\varepsilon\left(\mathbf{r}_j, \omega\right)\right] \mathrm{Tr}\left[\overline{\overline{\mathbf{G}}}\left(\mathbf{r}_i, \mathbf{r}_j, \omega\right) \overline{\overline{\mathbf{G}}}^{\dagger}\left(\mathbf{r}_i, \mathbf{r}_j, \omega\right)\right], \quad (2)$$

where $k_0 = \omega\sqrt{\mu_0 \varepsilon_0}$ is the magnitude of the vacuum wavevector in terms of the vacuum permeability $\mu_0$ and vacuum permittivity $\varepsilon_0$, $\Delta V_i$ is the volume of the $i$th particle, $\varepsilon(\mathbf{r}_i, \omega)$ is the dielectric function of the $i$th particle, $\mathbf{r}_i$ is the location vector for the center of mass of the $i$th particle, $\overline{\overline{\mathbf{G}}}\left(\mathbf{r}_i, \mathbf{r}_j, \omega\right)$ is the system Green's function (i.e., a dyadic defining electromagnetic interactions between particles $i$ and $j$), and $\dagger$ is the conjugate transpose operator.

In calculating the total power dissipated in each particle, the most challenging step is determining the system Green's function $\overline{\overline{\mathbf{G}}}\left(\mathbf{r}_i, \mathbf{r}_j, \omega\right)$. In the following sections, a method for calculating $\overline{\overline{\mathbf{G}}}\left(\mathbf{r}_i, \mathbf{r}_j, \omega\right)$ that may be applied to systems of spherical dipoles, nonspherical dipoles, and any combination thereof is presented. This method is a generalized many-body approach for modeling NFRHT and improves upon previous many-body approaches that have been restricted to spherical dipoles alone [3,17,30,31].

### B. System Green's function for *N* dipoles

For any collection of objects, the system Green's function $\overline{\overline{\mathbf{G}}}(\mathbf{r}, \mathbf{r}', \omega)$ may be defined solely in terms of the free-space Green's function $\overline{\overline{\mathbf{G}}}^0(\mathbf{r}, \mathbf{r}', \omega)$ and the electromagnetic and volumetric parameters of every object in the system [9,32]:

$$\overline{\overline{\mathbf{G}}}(\mathbf{r}, \mathbf{r}', \omega) = \overline{\overline{\mathbf{G}}}^0(\mathbf{r}, \mathbf{r}', \omega) + k_0^2 \int_V \overline{\overline{\mathbf{G}}}^0(\mathbf{r}, \mathbf{r}'', \omega) \varepsilon_{\mathrm{r}}(\mathbf{r}'', \omega) \overline{\overline{\mathbf{G}}}(\mathbf{r}'', \mathbf{r}', \omega)\, d^3\mathbf{r}'', \quad (3)$$

where $\mathbf{r}$ is the observation point, $\mathbf{r}'$ is the source point, $V$ is the volume of all objects and $\varepsilon_{\mathrm{r}}(\mathbf{r}'', \omega)$ is the relative dielectric function defined as the difference between the dielectric function of the objects and the nonabsorbing background reference medium, $\varepsilon_{\mathrm{r}}(\mathbf{r}'',\omega) = \varepsilon(\mathbf{r}'',\omega) - \varepsilon_{\mathrm{ref}}(\omega)$. The free-space Green's function $\overline{\overline{\mathbf{G}}}^{0}(\mathbf{r},\mathbf{r}',\omega)$ has the known analytical solution

$$\overline{\overline{\mathbf{G}}}^{0}(\mathbf{r},\mathbf{r}',\omega) = \left[\overline{\overline{\mathbf{I}}} + \frac{1}{k_{\mathrm{ref}}^{2}}\nabla\nabla\right] G^{0}(\mathbf{r},\mathbf{r}',\omega), \tag{4}$$

where $\overline{\overline{\mathbf{I}}}$ is the unit dyadic, $k_{\mathrm{ref}} = k_0\sqrt{\varepsilon_{\mathrm{ref}}(\omega)}$ is the wavevector magnitude in the nonabsorbing background reference medium, and the scalar free-space Green's function is defined for outgoing waves as

$$G^{0}(\mathbf{r},\mathbf{r}',\omega) = \frac{\mathrm{e}^{ik_{\mathrm{ref}}|\mathbf{r}-\mathbf{r}'|}}{4\pi|\mathbf{r}-\mathbf{r}'|}. \tag{5}$$

For $N$ dipoles, Eq. (3) takes on the form common to the many-body approach for NFRHT [3]:

$$\overline{\overline{\mathbf{G}}}(\mathbf{r}_i,\mathbf{r}_j,\omega) = \overline{\overline{\mathbf{G}}}^{0}(\mathbf{r}_i,\mathbf{r}_j,\omega) + k_0^2 \sum_{k=1}^{N} \Delta V_k \overline{\overline{\mathbf{G}}}^{0}(\mathbf{r}_i,\mathbf{r}_k,\omega)\varepsilon_{\mathrm{r}}(\mathbf{r}_k,\omega)\overline{\overline{\mathbf{G}}}(\mathbf{r}_k,\mathbf{r}_j,\omega). \tag{6}$$

The full system of equations given by Eq. (6) may be written in matrix form as

$$\begin{bmatrix} \overline{\overline{\mathbf{G}}}(\mathbf{r}_1,\mathbf{r}_1) & \cdots & \overline{\overline{\mathbf{G}}}(\mathbf{r}_1,\mathbf{r}_N) \\ \vdots & \ddots & \vdots \\ \overline{\overline{\mathbf{G}}}(\mathbf{r}_N,\mathbf{r}_1) & \cdots & \overline{\overline{\mathbf{G}}}(\mathbf{r}_N,\mathbf{r}_N) \end{bmatrix} = \left\{ \begin{bmatrix} \overline{\overline{\mathbf{I}}} & 0 & 0 \\ 0 & \ddots & 0 \\ 0 & 0 & \overline{\overline{\mathbf{I}}} \end{bmatrix} - k_0^2 \begin{bmatrix} \overline{\overline{\mathbf{G}}}^{0}(\mathbf{r}_1,\mathbf{r}_1) & \cdots & \overline{\overline{\mathbf{G}}}^{0}(\mathbf{r}_1,\mathbf{r}_N) \\ \vdots & \ddots & \vdots \\ \overline{\overline{\mathbf{G}}}^{0}(\mathbf{r}_N,\mathbf{r}_1) & \cdots & \overline{\overline{\mathbf{G}}}^{0}(\mathbf{r}_N,\mathbf{r}_N) \end{bmatrix} \right.$$

$$\left. \times \begin{bmatrix} \Delta V_1 \varepsilon_{\mathrm{r}}(\mathbf{r}_1,\omega) & 0 & 0 \\ 0 & \ddots & 0 \\ 0 & 0 & \Delta V_N \varepsilon_{\mathrm{r}}(\mathbf{r}_N,\omega) \end{bmatrix} \right\}^{-1} \begin{bmatrix} \overline{\overline{\mathbf{G}}}^{0}(\mathbf{r}_1,\mathbf{r}_1) & \cdots & \overline{\overline{\mathbf{G}}}^{0}(\mathbf{r}_1,\mathbf{r}_N) \\ \vdots & \ddots & \vdots \\ \overline{\overline{\mathbf{G}}}^{0}(\mathbf{r}_N,\mathbf{r}_1) & \cdots & \overline{\overline{\mathbf{G}}}^{0}(\mathbf{r}_N,\mathbf{r}_N) \end{bmatrix}. \tag{7}$$

In deriving Eq. (6), it is assumed that the system Green's function $\overline{\overline{\mathbf{G}}}(\mathbf{r},\mathbf{r}',\omega)$ is well-behaved for all $\mathbf{r}$ and $\mathbf{r}'$ [33] and may be approximated as constant over the domain of a given dipole. As such, the system Green's function over dipole domains may be represented by its value at dipole centers of mass,

$$\frac{1}{\Delta V_i \Delta V_j} \int_{\Delta V_i} \int_{\Delta V_j} \overline{\overline{\mathbf{G}}}(\mathbf{r},\mathbf{r}',\omega) d^3\mathbf{r}' d^3\mathbf{r} \approx \frac{1}{\Delta V_i \Delta V_j} \int_{\Delta V_i} \int_{\Delta V_j} \overline{\overline{\mathbf{G}}}(\mathbf{r}_i,\mathbf{r}_j,\omega) d^3\mathbf{r}' d^3\mathbf{r} = \overline{\overline{\mathbf{G}}}(\mathbf{r}_i,\mathbf{r}_j,\omega),$$

$$\text{for all } i \text{ and } j. \tag{8}$$

The free-space Green's function $\overline{\overline{\mathbf{G}}}^0(\mathbf{r},\mathbf{r}',\omega)$, on the other hand, must be handled with more care. In Eq. (6), the free-space Green's function takes two forms: one form when the observation and source points are in different dipoles ($i \neq j$), and one form when the observation and source points reside in the same dipole ($i = j$). When the observation and source points are within two different dipoles, it is acceptable that the free-space Green's function over the dipole domains be approximated by that at the dipole center-of-mass locations,

$$\frac{1}{\Delta V_i \Delta V_j}\int_{\Delta V_i}\int_{\Delta V_j}\overline{\overline{\mathbf{G}}}^0(\mathbf{r},\mathbf{r}',\omega)d^3\mathbf{r}'d^3\mathbf{r} \approx \frac{1}{\Delta V_i \Delta V_j}\int_{\Delta V_i}\int_{\Delta V_j}\overline{\overline{\mathbf{G}}}^0\left(\mathbf{r}_i,\mathbf{r}_j,\omega\right)d^3\mathbf{r}'d^3\mathbf{r}$$

$$= \overline{\overline{\mathbf{G}}}^0\left(\mathbf{r}_i,\mathbf{r}_j,\omega\right), \text{ for } i \neq j. \tag{9}$$

Here, the free-space Green's function in Cartesian coordinates is given by [33,34]:

$$\overline{\overline{\mathbf{G}}}^0\left(\mathbf{r}_i,\mathbf{r}_j,\omega\right) = \frac{\exp\left(ik_{\text{ref}}r_{ij}\right)}{4\pi r_{ij}}\left[\left(1 - \frac{1}{\left(k_{\text{ref}}\, r_{ij}\right)^2} + \frac{i}{k_{\text{ref}}\, r_{ij}}\right)\overline{\overline{\mathbf{I}}}\right.$$

$$\left. - \left(1 - \frac{3}{(k_{\text{ref}}\, r_{ij})^2} + \frac{3i}{k_{\text{ref}}\, r_{ij}}\right)\left(\hat{\mathbf{r}}_{ij}\hat{\mathbf{r}}_{ij}^{\dagger}\right)\right] \text{ for } i \neq j, \tag{10}$$

where $r_{ij} = \left|\mathbf{r}_i - \mathbf{r}_j\right|$ and $\hat{\mathbf{r}}_{ij} = \frac{(\mathbf{r}_i - \mathbf{r}_j)}{|\mathbf{r}_i - \mathbf{r}_j|}$.

Up until this point, the geometry of dipoles has been irrelevant. When the observation and source points are within the same dipole domain, however, $\overline{\overline{\mathbf{G}}}^0(\mathbf{r}_i,\mathbf{r}_i,\omega)$ becomes symbolic in nature and represents explicit integration over the geometry of the dipole domain,

$$\overline{\overline{\mathbf{G}}}^0(\mathbf{r}_i,\mathbf{r}_i,\omega) \equiv \frac{1}{\Delta V_i}\int_{\Delta V_i}\overline{\overline{\mathbf{G}}}^0(\mathbf{r}_i,\mathbf{r}',\omega)\, d^3\mathbf{r}'. \tag{11}$$

The integral in Eq. (11) remains because there is a singularity that exists in the free-space Green's function when $\mathbf{r}' = \mathbf{r}_i$. As such, the free-space Green's function for dipole self-interaction cannot be represented by its value at the dipole center of mass, and integrals must be solved using principal value techniques [35-38].

The limits of integration in Eq. (11) are a function of dipole geometry and orientation. This geometric dependence leads to one of the main points of this paper: to incorporate nonspherical dipole geometries into many-body theories of NFRHT, all one must do is include the proper limits of integration in Eq. (11).

### C. Principal value solution of the free-space Green's function for dipole self-interaction

The free-space Green's function for dipole self-interaction may be written in terms of its principal value ($\bar{\bar{\mathbf{M}}}_i$ term) and depolarization factor ($\bar{\bar{\mathbf{L}}}_i$ term) [33,36,39]:

$$\bar{\bar{\mathbf{G}}}^0(\mathbf{r}_i, \mathbf{r}_i, \omega) = \frac{1}{\Delta V_i}\left[\bar{\bar{\mathbf{M}}}_i - \frac{\bar{\bar{\mathbf{L}}}_i}{k_{\mathrm{ref}}^2}\right]. \tag{12}$$

The $\bar{\bar{\mathbf{M}}}_i$ dyadic is defined as

$$\bar{\bar{\mathbf{M}}}_i = \lim_{\delta \to 0} \int_{\Delta V_i - V_\delta} \bar{\bar{\mathbf{G}}}^0(\mathbf{r}_i, \mathbf{r}', \omega) d^3\mathbf{r}', \tag{13}$$

where $V_\delta$ is an exclusion volume of vanishing dimension $\delta$ around $\mathbf{r}' = \mathbf{r}_i$.

Physically, $\bar{\bar{\mathbf{L}}}_i$ may be interpreted as a depolarizing factor for the exclusion volume $V_\delta$ and is expressed mathematically as the surface integral

$$\bar{\bar{\mathbf{L}}}_i = \frac{1}{4\pi} \int_{S_\delta} \frac{\hat{\mathbf{n}}(\mathbf{r}')\,(\hat{\mathbf{e}}_R)^{\mathrm{T}}}{|\mathbf{r}_i - \mathbf{r}'|^2} dS', \tag{14}$$

where $S_\delta$ is the surface of the exclusion volume, $\hat{\mathbf{n}}(\mathbf{r}')$ is the unit vector normal to surface $S_\delta$, $\hat{\mathbf{e}}_R = \frac{\mathbf{r}' - \mathbf{r}_i}{|\mathbf{r}_i - \mathbf{r}'|}$ is the unit vector from the center of mass of the $i$th dipole to the surface of the exclusion volume, and T is the transpose operator. It is interesting to note that $\bar{\bar{\mathbf{L}}}_i$ depends only on the geometry of the exclusion volume and is a real, symmetric dyadic with unity trace [36].

*1. Analytical solution for spherical dipoles*

For spherical dipoles with spherical exclusion volumes, both the $\bar{\bar{\mathbf{M}}}_i$ and $\bar{\bar{\mathbf{L}}}_i$ dyadics may be derived analytically as $\bar{\bar{\mathbf{M}}}_i = \frac{2}{3}\frac{\bar{\bar{\mathbf{I}}}}{k_{\text{ref}}^2}\left[e^{ik_{\text{ref}}R_{s,i}}\left(1 - ik_{\text{ref}}R_{s,i}\right) - 1\right]$ and $\bar{\bar{\mathbf{L}}}_i = \bar{\bar{\mathbf{I}}}/3$ [36,38,40]. In this case, Eq. (12) reduces to

$$\bar{\bar{\mathbf{G}}}^0(\mathbf{r}_i, \mathbf{r}_i, \omega) = \frac{1}{\Delta V_i}\frac{\bar{\bar{\mathbf{I}}}}{k_{\text{ref}}^2}\left\{\frac{2}{3}e^{ik_{\text{ref}}R_{s,i}}\left(1 - ik_{\text{ref}}R_{s,i}\right) - 1\right\}, \tag{15}$$

where $R_{s,i}$ is the radius of the $i$th spherical dipole.

*2. Numerical solution for any nonspherical dipoles*

For nonspherical dipoles of arbitrary orientation, it is often desirable to solve the integral in Eq. (11) numerically. To implement a numerical solver with a finite exclusion volume $V_{d,i}$ around the singularity point, the integral in Eq. (11) may be decomposed into two parts: one integral with no singularity point over the region $\Delta V_i - V_{d,i}$ and one integral over the region $V_{d,i}$ that contains the singularity [36]:

$$\bar{\bar{\mathbf{G}}}^0(\mathbf{r}_i, \mathbf{r}_i, \omega) = \frac{1}{\Delta V_i}\left[\int_{\Delta V_i - V_{d,i}} \bar{\bar{\mathbf{G}}}^0(\mathbf{r}_i, \mathbf{r}', \omega) d^3\mathbf{r}' + \int_{V_{d,i}} \bar{\bar{\mathbf{G}}}^0(\mathbf{r}_i, \mathbf{r}', \omega) d^3\mathbf{r}'\right]. \tag{16}$$

Implementing the principal value technique for the second integral, Eq. (16) becomes

$$\bar{\bar{\mathbf{G}}}^0(\mathbf{r}_i, \mathbf{r}_i, \omega) = \frac{1}{\Delta V_i}\left[\int_{\Delta V_i - V_{d,i}} \bar{\bar{\mathbf{G}}}^0(\mathbf{r}_i, \mathbf{r}', \omega) d^3\mathbf{r}' + \lim_{\delta \to 0}\int_{V_{d,i} - V_\delta} \bar{\bar{\mathbf{G}}}^0(\mathbf{r}_i, \mathbf{r}', \omega) d^3\mathbf{r}' - \frac{\bar{\bar{\mathbf{L}}}_i}{k_{\text{ref}}^2}\right], \tag{17}$$

where the $\bar{\bar{\mathbf{M}}}_i$ dyadic corresponds to the sum of the first two terms inside the square brackets. From here, it is computationally desirable to define a spherical $V_{d,i}$ and a spherical $V_\delta$ so that the last two terms of Eq. (17) may be reduced to their analytical forms. If spherical $V_{d,i}$ and $V_\delta$ domains are chosen, Eq. (17) may then be written succinctly as

$$\bar{\bar{\mathbf{G}}}^0(\mathbf{r}_i, \mathbf{r}_i, \omega) = \frac{1}{\Delta V_i}\left[\int_{\Delta V_i - V_{d,i}} \bar{\bar{\mathbf{G}}}^0(\mathbf{r}_i, \mathbf{r}', \omega) d^3\mathbf{r}' + \frac{\bar{\bar{\mathbf{I}}}}{k_{\text{ref}}^2}\left\{\frac{2}{3}e^{ik_{\text{ref}}R_{d,i}}\left(1 - ik_{\text{ref}}R_{d,i}\right) - 1\right\}\right], \tag{18}$$

where $R_{d,i}$ is the radius of the finite spherical exclusion volume $V_{d,i}$ for the $i$th dipole. Eq. (18) is the final form of the free-space Green's function for dipole self-interaction that is implemented in the generalized many-body method used to generate results in this paper. The integral over $\Delta V_i - V_{d,i}$ is where nonspherical dipole geometries come into play, and this integral can be calculated numerically using standard integral solvers.

*3. Numerical solution for ellipsoidal dipoles of arbitrary orientation*

To define the upper limit of integration over $\Delta V_i$ in Eq. (18), the volume integral may be expanded as a triple integral over spherical coordinates, $\int_{\Delta V_i - V_{d,i}} d^3\mathbf{r} \rightarrow$ $\int_0^{2\pi} \int_0^{\pi} \int_{R_{\min}(\phi,\theta)}^{R_{\max}(\phi,\theta)} R^2 \sin\phi \, dR \, d\phi \, d\theta$, where $R$ is the radial coordinate, $\phi$ is the polar angle, $\theta$ is the azimuthal angle, $R_{\min}(\phi,\theta)$ is the radial expression for the bounding surface of the finite exclusion volume $V_{d,i}$, and $R_{\max}(\phi,\theta)$ is the radial expression for the bounding surface of the dipole volume $\Delta V_i$. This formulation is useful since many common geometric surfaces can be expressed analytically in terms of spherical coordinates. For example, if a spherical finite exclusion volume is used in Eq. (18), then $R_{\min}(\phi,\theta) = R_{s,i}$.

Ellipsoidal surfaces defined by three unique semiaxes $a$, $b$, and $c$, may be expressed analytically as functions of $R$, $\theta$, and $\phi$. Due to the wide range of geometries that can be constructed by changing the three semiaxis dimensions, ellipsoidal dipoles are good candidates to study the effect of geometry on NFRHT in systems of many particles. An expression for the radial component $R(\phi,\theta)$ of a rotated ellipsoid may be found by solving the general equation for a rotated ellipsoid. The found expression for $R(\phi,\theta)$ may then be applied as the upper limit of integration in Eq. (18) to calculate NFRHT between ellipsoidal dipoles of variable dimension and orientation.

The general equation for a rotated ellipsoidal surface may be written as the matrix product

$$\mathbf{x}^{\mathrm{T}}\mathbf{R}^{\mathrm{T}}\mathbf{A}\,\mathbf{R}\,\mathbf{x} = 1. \tag{19}$$

Here, $\mathbf{x}$ is a general location vector with components written in spherical coordinates,

$$\mathbf{x} = \begin{bmatrix} x \\ y \\ z \end{bmatrix} = \begin{bmatrix} R(\phi,\theta)\cos\theta\sin\phi \\ R(\phi,\theta)\sin\theta\sin\phi \\ R(\phi,\theta)\cos\phi \end{bmatrix}, \tag{20}$$

$\mathbf{R}$ is a rotation matrix composed of consecutive rotations $\theta_x$, $\theta_y$, and $\theta_z$ around the $x$-, $y$-, and $z$-axes, respectively,

$$\mathbf{R} = \mathbf{R}_x(\theta_x)\mathbf{R}_y(\theta_y)\mathbf{R}_z(\theta_z)$$

$$= \begin{bmatrix} 1 & 0 & 0 \\ 0 & \cos\theta_x & -\sin\theta_x \\ 0 & \sin\theta_x & \cos\theta_x \end{bmatrix} \begin{bmatrix} \cos\theta_y & 0 & \sin\theta_y \\ 0 & 1 & 0 \\ -\sin\theta_y & 0 & \cos\theta_y \end{bmatrix} \begin{bmatrix} \cos\theta_z & -\sin\theta_z & 0 \\ \sin\theta_z & \cos\theta_z & 0 \\ 0 & 0 & 1 \end{bmatrix}, \tag{21}$$

and $\mathbf{A}$ is a 3-by-3 diagonal matrix of the squared inverses of the ellipsoid semiaxes,

$$\mathbf{A} = \begin{bmatrix} 1/a^2 & 0 & 0 \\ 0 & 1/b^2 & 0 \\ 0 & 0 & 1/c^2 \end{bmatrix}. \tag{22}$$

Semiaxes $a$, $b$, and $c$ are aligned, respectively, with the $x$-, $y$-, and $z$-axes of the local coordinate system of the ellipsoid (see Fig. 2).

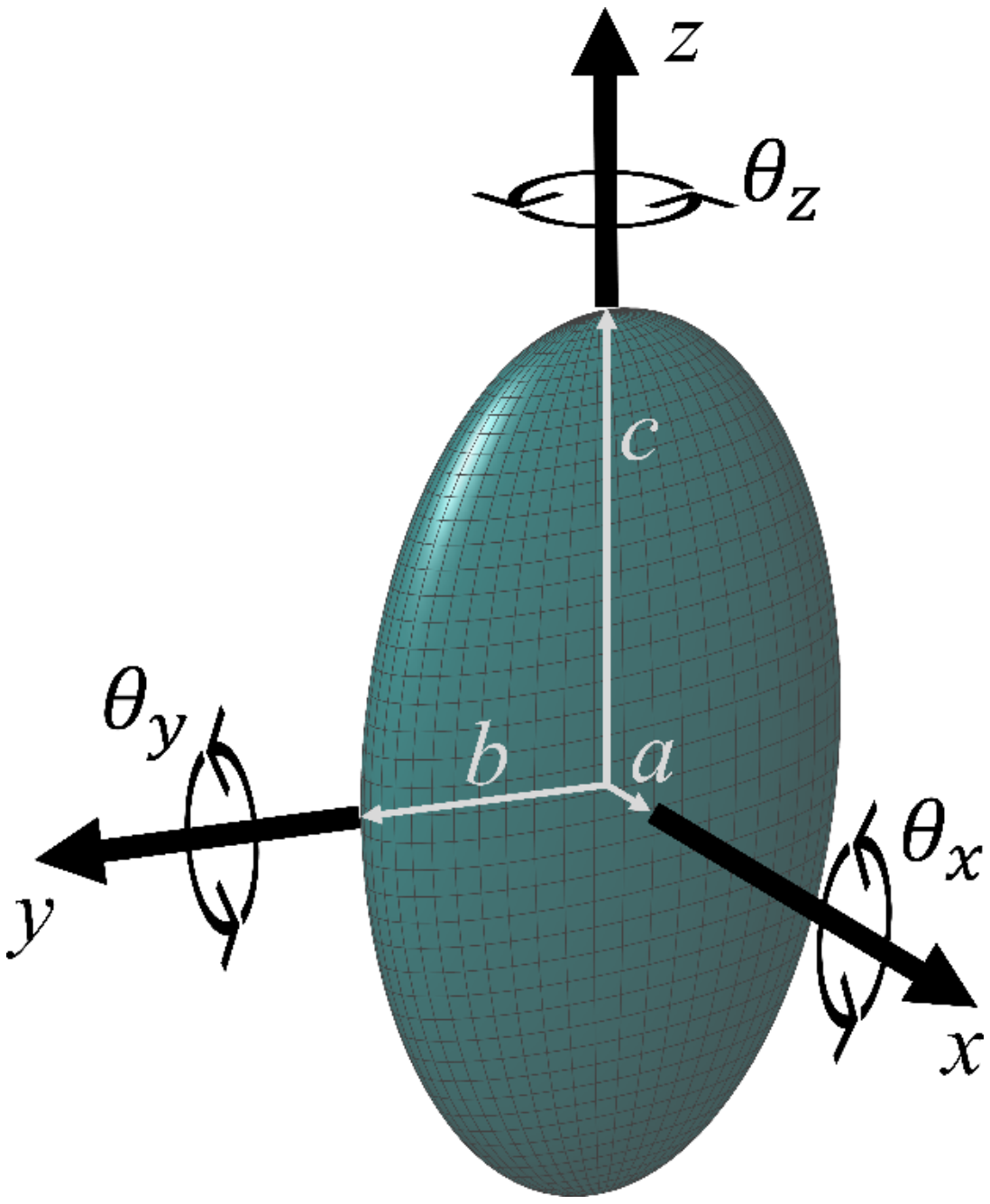

FIG. 2. Schematic of ellipsoidal particle with semiaxis lengths $a$, $b$, and $c$. Rotations are applied with respect to the local Cartesian coordinate system of the ellipsoid that is aligned with the three ellipsoid semiaxes.

Once all semiaxis values $a$, $b$, and $c$ and rotation angles $\theta_x$, $\theta_y$, and $\theta_z$ are input into Eq. (19), this matrix equation may be solved to find an expression $R(\phi, \theta)$ for the rotated ellipsoidal surface.

## D. Strong and weak forms

The generalized many-body approach for calculating NFRHT presented thus far is of the strong form. In the strong form, Eq. (11) is solved in full using the principal value method such that both the $\bar{\bar{\mathbf{M}}}_i$ and $\bar{\bar{\mathbf{L}}}_i$ dyadics are included in calculating the free-space Green's function for dipole self-interaction.

In the limit of very small particles, however, the exclusion volume $V_\delta$ used in the principal value method may be approximated to be the same as the dipole domain (i.e., $V_\delta \approx \Delta V_i$) [33,39]. In this case, the $\bar{\bar{\mathbf{M}}}_i$ dyadic approaches zero, and the free-space Green's function for dipole self-interaction may be expressed in its weak form as

$$\bar{\bar{\mathbf{G}}}^0_{\mathrm{weak}}(\mathbf{r}_i, \mathbf{r}_i, \omega) = -\frac{\bar{\bar{\mathbf{L}}}_i}{\Delta V_i\, k_{\mathrm{ref}}^2}. \tag{23}$$

Since the weak form presented in Eq. (23) is derived as the limiting case when the outer dipole domain $\Delta V_i$ shrinks down to the length scale of the exclusion volume $V_\delta$, the $\bar{\bar{\mathbf{L}}}_i$ dyadic must retain the geometric parameters of the dipole volume $\Delta V_i$, not the geometric parameters of the arbitrarily shaped exclusion volume $V_\delta$ used in the strong form. As such, starting with the strong form given by Eq. (12) and then setting $\bar{\bar{\mathbf{M}}}_i = 0$ does not always result in the proper weak form if the strong form was calculated using dipoles and exclusion volumes of different geometries.

For ellipsoidal dipoles, the $\bar{\bar{\mathbf{L}}}_i$ dyadic used in the weak form given by Eq. (23) may be solved analytically as [36,41]

$$\bar{\bar{\mathbf{L}}}_{i,\mathrm{ellipsoid}} = \begin{bmatrix} L_{i,1} & 0 & 0 \\ 0 & L_{i,2} & 0 \\ 0 & 0 & L_{i,3} \end{bmatrix}, \tag{24}$$

with components expanded as

$$L_{i,1} = \frac{abc}{2}\int_0^\infty (a^2+q)^{-1}[(q+a^2)(q+b^2)(q+c^2)]^{-1/2} dq, \tag{25}$$

$$L_{i,2} = \frac{abc}{2}\int_0^\infty (b^2+q)^{-1}[(q+a^2)(q+b^2)(q+c^2)]^{-1/2} dq, \tag{26}$$

$$L_{i,3} = \frac{abc}{2}\int_0^\infty (c^2+q)^{-1}[(q+a^2)(q+b^2)(q+c^2)]^{-1/2} dq. \tag{27}$$

Rotation of ellipsoidal dipoles in the global coordinate system may be applied as $\mathbf{R}^{\mathrm{T}}\bar{\bar{\mathbf{L}}}_{i,\mathrm{ellipsoid}}\mathbf{R}$.

Usually, numerical calculation of the strong form of the self-term of the free-space Green's function given by Eq. (18) is much more time consuming than using a purely analytical solution for the weak form. The computational complexity arises in numerical calculation of the integral in Eq. (18) and depends on the type of integral solver that is chosen. High-precision numerical integration is required to account for the strong variation in the free-space Green's function at small values of $|\mathbf{r}' - \mathbf{r}_i|$. For instance, using the MATLAB `vpaintegral` function on a standard desktop computer, it took ~32 seconds to calculate the self-term of the free-space Green's function for an ellipsoid ($a = 15$ nm, $b = 45$ nm, $c = 75$ nm) using the strong form [Eq. (18)] and ~0.06 seconds using the weak form [Eq. (23)]. In these strong form calculations, the radius of the finite spherical exclusion volume was set as $V_{a,i} = a/2$ to achieve numerical convergence [36]. While more efficient implementations could improve on the calculation times presented here, calculation of analytical weak forms is always more efficient than strong forms requiring numerical integration.

Finally, it is important to note that the weak form expressed in Eq. (23) does not satisfy energy conservation or the optical theorem [33]. As such, NFRHT results derived using the weak form are not true physical solutions, and care must be taken to ensure that weak approximations are appropriate for a given system. One method for checking the applicability of strong and weak forms is through calculation of particle polarizability, as discussed in the next section.

## III. POLARIZABILITY

The generalized many-body approach developed in this paper does not require the definition of dipole polarizability. However, an equation for the dipole polarizability may be derived *a posteriori* for integration into other existing models of NFRHT. Additionally, the dipole polarizability is a useful metric to assess the limit of applicability of the weak form of the generalized many-body framework presented here. Expressions for and analysis of the polarizability of spherical and ellipsoidal dipoles is provided hereafter.

### A. Definition

Previously, the self-interaction of the free-space Green's function was defined in terms of the dyadics $\overline{\overline{\mathbf{M}}}_i$ and $\overline{\overline{\mathbf{L}}}_i$. Introducing these dyadics is useful because these two quantities may be used to define a general expression for the dipole polarizability tensor and distinguish between strong and weak forms of polarizability that commonly arises in other models of NFRHT. The dipole polarizability tensor defined in terms of $\overline{\overline{\mathbf{M}}}_i$ and $\overline{\overline{\mathbf{L}}}_i$ is [33]:

$$\overline{\overline{\boldsymbol{\alpha}}}_i = \Delta V_i \varepsilon_0 \varepsilon_{\mathrm{r}}(\mathbf{r}_i,\omega)\left\{\overline{\overline{\mathbf{I}}} - \left[k_0^2 \overline{\overline{\mathbf{M}}}_i - \frac{\overline{\overline{\mathbf{L}}}_i}{\varepsilon_{\mathrm{ref}}(\omega)}\right]\varepsilon_{\mathrm{r}}(\mathbf{r}_i,\omega)\right\}^{-1}. \tag{28}$$

Written explicitly in terms of the free-space Green's function, Eq. (28) becomes

$$\overline{\overline{\boldsymbol{\alpha}}}_i = \varepsilon_0 \varepsilon_{\mathrm{r}}(\mathbf{r}_i,\omega)\left[\frac{\overline{\overline{\mathbf{I}}}}{\Delta V_i} - k_0^2 \overline{\overline{\mathbf{G}}}^0(\mathbf{r}_i,\mathbf{r}_i,\omega)\varepsilon_{\mathrm{r}}(\mathbf{r}_i,\omega)\right]^{-1}. \tag{29}$$

For spherical dipoles, the diagonal elements of the $\overline{\overline{\mathbf{L}}}_i$ dyadic are equal to 1/3 such that the weak form of the polarizability reduces to the electrostatic polarizability (i.e., the Clausius-Mossotti polarizability) defined as [33,41,42]

$$\overline{\overline{\boldsymbol{\alpha}}}_i^{(\mathrm{weak,sphere})} = \overline{\overline{\boldsymbol{\alpha}}}_i^{(\mathrm{CM})} = 3\varepsilon_0 \varepsilon_{\mathrm{ref}}(\omega)\Delta V_i[\varepsilon(\mathbf{r}_i,\omega) - \varepsilon_{\mathrm{ref}}][\varepsilon(\mathbf{r}_i,\omega) + 2\varepsilon_{\mathrm{ref}}]^{-1}\overline{\overline{\mathbf{I}}}. \tag{30}$$

Some researchers have also modified the weak form of the polarizability for spherical dipoles to include correction factors for radiation damping. For example, a standard expression for the polarizability used in many-body approaches for spherical dipoles is [15,17,43]:

$$\overline{\overline{\boldsymbol{\alpha}}}_i^{(\mathrm{weak,sphere,corrected})} = \varepsilon_0 \left(\frac{1}{\alpha_i^{(\mathrm{CM})}} - \frac{i k_{\mathrm{ref}}^3}{6\pi\varepsilon_{\mathrm{ref}}}\right)^{-1}\overline{\overline{\mathbf{I}}}. \tag{31}$$

Eq. (31) was originally derived by Draine [44] for light scattering by particles small compared to the wavelength (i.e., $k_0 R_{s,i} < 1$). It is crucial to note that while the radiative correction may provide a more accurate solution than the uncorrected weak form of the polarizability tensor, it is not equivalent to the strong form of polarizability given independently by Eqs. (28) and (29). A thorough discussion of strong versus weak forms of dipole polarizability is given by Lakhtakia in Ref. [39].

For ellipsoidal dipoles, the weak form of the polarizability tensor is obtained using Eq. (28) with $\overline{\overline{\mathbf{M}}}_i = 0$ and $\overline{\overline{\mathbf{L}}}_i$ calculated from Eq. (24)

$$\overline{\overline{\boldsymbol{\alpha}}}_i^{(\text{weak,ellipsoid})} = \varepsilon_0 \alpha_i^{(0)} \left\{ \overline{\overline{\mathbf{I}}} + \frac{\overline{\overline{\mathbf{L}}}_i}{\varepsilon_{\text{ref}}(\omega)} \frac{\alpha_i^{(0)}}{\Delta V_i} \right\}^{-1}. \quad (32)$$

Nikbakht [25] has modified this weak form of the polarizability for ellipsoidal dipoles to include a radiative correction

$$\overline{\overline{\boldsymbol{\alpha}}}_i^{(\text{weak,ellipsoid,corrected})} = \varepsilon_0 \overline{\overline{\boldsymbol{\alpha}}}_i^{(\text{weak,ellipsoid})} \left\{ \overline{\overline{\mathbf{I}}} - \frac{i k_{\text{ref}}}{6\pi\varepsilon_{\text{ref}}} \overline{\overline{\boldsymbol{\alpha}}}_i^{(\text{weak,ellipsoid})} \right\}^{-1}. \quad (33)$$

**B. Application to other NFRHT models**

To model nonspherical dipoles using existing dipole formalisms of NFRHT, all that must be done is to insert Eq. (28) as the polarizability or apply Eq. (11) for the self-interaction terms of the free-space Green's function.

In exciting field formalisms, such as the original many-body approach developed by Ben-Abdallah *et al*. [3], the general polarizability tensor given by Eq. (28) should be inserted into the system of equations as:

$$\mathbf{p}_i^{(\text{ind})} = \overline{\overline{\boldsymbol{\alpha}}}_i \frac{k_{\text{ref}}^2}{\varepsilon_{\text{ref}}} \sum_{j \neq i} \overline{\overline{\mathbf{G}}}^0\left(\mathbf{r}_i, \mathbf{r}_j, \omega\right) \cdot \mathbf{p}_j, \quad (34)$$

where the induced dipole moment $\mathbf{p}_i^{(\mathrm{ind})}$ is related to the external exciting electric field $\mathbf{E}_i^{(\mathrm{exc})}$ as $\mathbf{p}_i^{(\mathrm{ind})} = \overline{\overline{\boldsymbol{\alpha}}}_i \mathbf{E}_i^{(\mathrm{exc})}$, and $\mathbf{p}_j$ is the total dipole moment of the $j$th dipole.

In actual field formulations, such as the thermal discrete dipole approximation developed by Edalatpour *et al*. [45], the self-interaction terms of the free-space Green's function given by Eq. (18) should be used in the system of equations:

$$\mathbf{p}_i^{(\mathrm{ind})} = \alpha_i^{(0)} \frac{k_{\mathrm{ref}}^2}{\varepsilon_{\mathrm{ref}}} \sum_j \overline{\overline{\mathbf{G}}}^0\left(\mathbf{r}_i, \mathbf{r}_j, \omega\right) \cdot \mathbf{p}_j, \tag{35}$$

where the induced dipole moment $\mathbf{p}_i^{(\mathrm{ind})}$ is related to the actual electric field in the $i$th dipole $\mathbf{E}_i^{(\mathrm{act})}$ as $\mathbf{p}_i^{(\mathrm{ind})} = \alpha_i^{(0)} \mathbf{E}_i^{(\mathrm{act})}$. Here, $\alpha_i^{(0)}$ is the bare polarizability of the $i$th dipole, $\alpha_i^{(0)} = \varepsilon_0 \Delta V_i \varepsilon_{\mathrm{r}}(\mathbf{r}_i, \omega)$. More information on actual versus exciting field formulations for NFRHT may be found in Ref. [18].

### C. Comparison of strong versus weak forms of the polarizability of ellipsoidal dipoles

To better understand the differences between the strong and weak forms of polarizability, the polarizability of ellipsoidal dipoles made of SiC embedded in vacuum (i.e., $\varepsilon_{\mathrm{ref}} = 1$) is analyzed (Fig. 3). The dielectric function of SiC is calculated using a Lorentz oscillator model [46]:

$$\varepsilon(\omega) = \varepsilon_\infty \left(\frac{\omega^2 - \omega_{\mathrm{LO}}^2 + i\Gamma\omega}{\omega^2 - \omega_{\mathrm{TO}}^2 + i\Gamma\omega}\right), \tag{36}$$

where $\varepsilon_\infty = 6.7$, $\omega_{\mathrm{LO}} = 1.825 \times 10^{14}$ rad/s, $\omega_{\mathrm{TO}} = 1.494 \times 10^{14}$ rad/s, and $\Gamma = 8.966 \times 10^{11}$ rad/s.

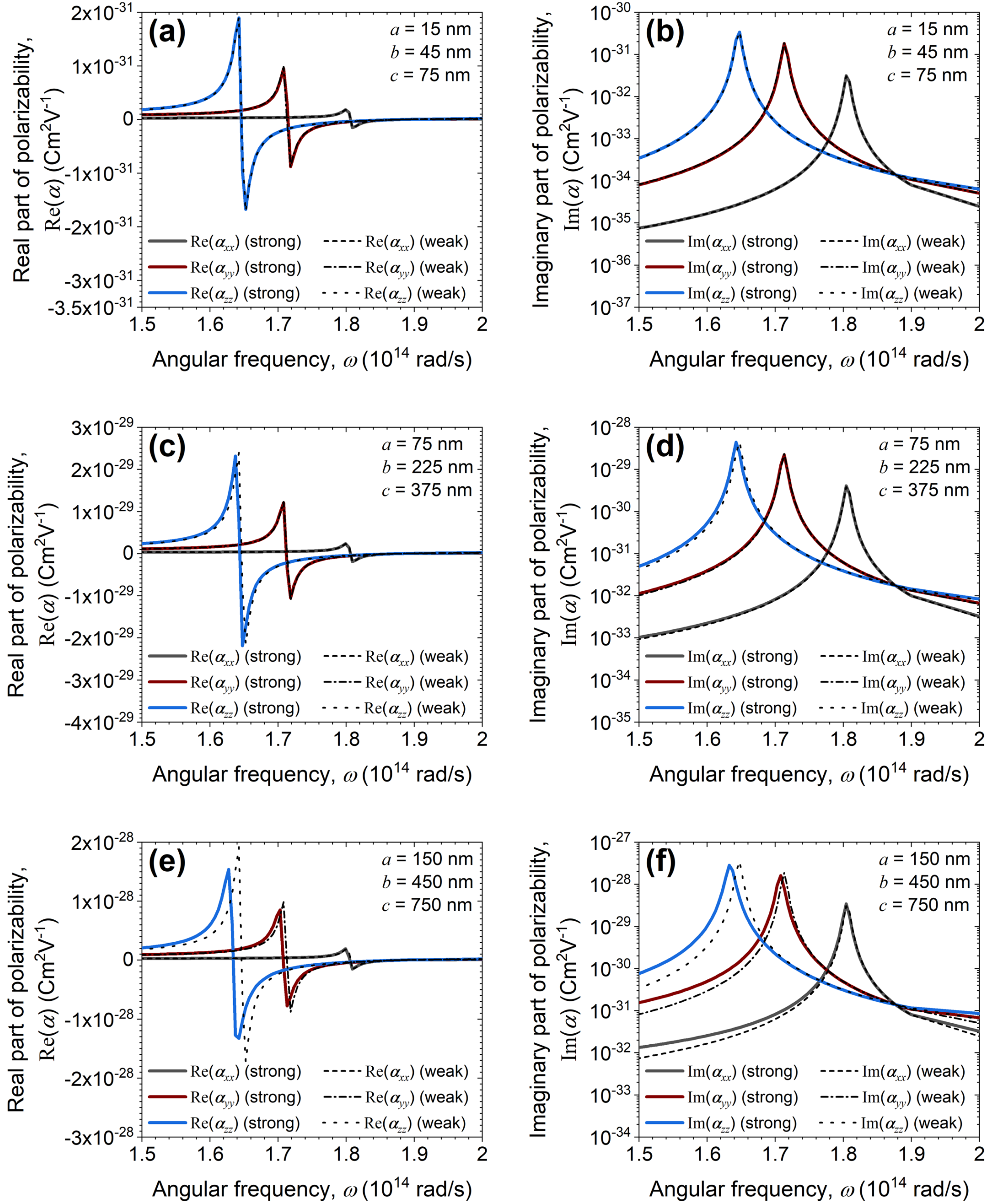

FIG. 3. Comparison of the strong and weak forms of polarizability for ellipsoidal dipoles. Three ellipsoidal dipole geometries are modeled: (a), (b) dimensions $a = 15$ nm, $b = 45$ nm, $c = 75$ nm and size parameter $X = 0.047$; (c), (d) dimensions $a = 75$ nm, $b = 225$ nm, $c = 375$ nm and size parameter $X = 0.24$; and (e), (f) dimensions $a = 150$ nm, $b = 450$ nm, $c = 750$ nm and size parameter $X = 0.47$. The ellipsoidal dipoles are made of SiC and are embedded in vacuum ($\varepsilon_{\text{ref}} = 1$).

The strong form of the polarizability is calculated from Eq. (29), with the self-interaction free-space Green's function found using Eq. (18). The upper limit of integration implemented in Eq. (18) is given by Eqs. (19)–(22), and the lower limit of integration is defined by the finite spherical exclusion volume of radius $R_{d,i} = \min\{a, b, c\}/2$. The strong and weak forms of the polarizability are calculated for three ellipsoidal dipole sizes: $a_1 = 15$ nm, $b_1 = 45$ nm, $c_1 = 75$ nm; $a_2 = 75$ nm, $b_2 = 225$ nm, $c_2 = 375$ nm; $a_3 = 150$ nm, $b_3 = 450$ nm, $c_3 = 750$ nm. These three cases correspond, respectively, to size parameters of 0.047, 0.24 and 0.47 calculated as $X = 2\pi L_{ch}/\lambda_T$ [41], where the thermal photon wavelength $\lambda_T$ is taken as 10 μm and the characteristic length is determined using $L_{\mathrm{ch}} = \max\{a, b, c\}$. It should be noted that particles are electrically small in all cases and thus in the dipole regime.

The strong and weak forms of the polarizability show good agreement for smaller ellipsoidal dipoles ($a = 15$ nm, $b = 45$ nm, $c = 75$ nm, $X = 0.047$) [Fig. 3(a)–(b)]. In this case, the weak form of the polarizability may be implemented in NFRHT calculations without loss of accuracy. For the larger ellipsoidal dipoles ($a = 150$ nm, $b = 450$ nm, $c = 750$ nm, $X = 0.47$) [Fig. 3(e)–(f)], however, the strong and weak forms of the polarizability show noticeable differences. In general, resonances in the weak form of the polarizability are blueshifted for these larger dipoles, with the most severe blueshift occurring for the resonance associated with the largest ellipsoid semiaxis. Therefore, when modeling NFRHT in ellipsoidal SiC dipoles of this size, the strong form of polarizability must be used to maintain accuracy. Polarizability approximations that are based on the weak form, such as those presented in Ref. [26], may lead to increased error in NFRHT calculations for larger particles. According to the examples modeled in Fig. 3, the strong form should be used in general for SiC particles with size parameter greater than ~0.24. This threshold is further confirmed in Sec. IV.C, after verification of the generalized many-body approach, where strong and weak calculations of NFRHT between two ellipsoidal SiC particles are compared. Below a size parameter of ~0.24, the weak form provides sufficiently accurate results.

## IV. RADIATIVE HEAT TRANSFER RESULTS AND DISCUSSION

The generalized many-body approach developed in Sec. II is applied to study NFRHT in three different systems. First, the solution for the total power dissipated in two SiC spheroids of variable orientation is compared with published results. Next, the generalized many-body theory is used to calculate the spectral conductance between two SiC ellipsoids of variable orientation. Third, strong- and weak-form NFRHT calculations between two SiC ellipsoidal particles are compared. Finally, the generalized many-body method is applied to calculate the spectral radiative thermal conductivity of a three-dimensional metamaterial composed of $SiO_2$ ellipsoidal particles. In these studies, SiC and $SiO_2$ are chosen because these materials support geometrically dependent LSPhs that dominate NFRHT in the infrared. All particles are embedded in vacuum (i.e., $\varepsilon_{\mathrm{ref}} = 1$), and the strong form of the generalized many-body approach is implemented throughout unless otherwise stated.

### A. Comparison against solution for two spheroidal dipoles

The generalized many-body method for spherical dipoles takes the exact same mathematical form as the discrete system Green's function method developed in Ref. [9] when each object is represented by a single subvolume. In Ref. [9], the discrete system Green's function method is verified with good agreement against the analytical solution for spheres and against other spherical dipole models, so the generalized many-body method is expected to perform accurately in calculating NFRHT between spherical dipoles. As such, the remaining verification check left to perform is for implementation of nonspherical geometries.

Towards this aim, we compare the generalized many-body method against an analytical NFRHT model for two spheroidal dipoles (i.e., dipoles with semiaxes $a = b \neq c$) presented by Incardone *et al*. [24]. Specifically, we use Eq. (4) in Ref. [24] to calculate the spectral and total power dissipated as one dipole is rotated by angles $0 \leq \theta_y \leq \pi/2$. (After contacting the authors of Ref. [24] , it was noted that Eq. (4) of Ref. [24] should be corrected to have squared rather than linear trigonometric functions. This corrected form of the analytical solution was implemented in this manuscript.) For the generalized many-body results, we use Eq. (18) to calculate the self-interaction terms of the free-space Green's function, where the spheroidal surfaces are defined by Eqs. (19)–(22) and the finite spherical exclusion volume $V_{d,i}$ has radius $R_{d,i} = \min\{a, b, c\}/2$. The spheroidal dipoles are of dimension $a = b = 15$ nm and $c = 75$ nm, arranged in parallel, and at a center-of-mass separation distance $d = 7L_{\text{ch}} = 525$ nm. One dipole is at temperature $T_1 = 0$ K and the other is at $T_2 = 300$ K.

The generalized many-body solution and analytical solution for the total power dissipated in a dipole match well (Fig. 4). The relative difference between these two solutions for total power dissipated in a dipole is under 2% across all orientations. This slight discrepancy between results may be attributed to differences in the way reflections are treated in each method: the analytical solution makes a single-reflection approximation, whereas the generalized many-body method accounts for multiple reflections. The spectra of power dissipated for the generalized many-body and analytical solutions are also in good agreement. Sample spectra are shown for $\theta_y = 0$ [Fig. 5(a)], $\theta_y = 0.25\pi$ [Fig. 5(b)], and $\theta_y = 0.47\pi$ [Fig. 5(c)]. The relative difference between the spectra of the generalized many-body and analytical solutions is the largest (~8%) for all three $\theta_y$ values at a frequency of $10^{14}$ rad/s where the spectral power dissipated has no perceptible impact on the total power dissipated. Based on this comparison with the analytical solution of NFRHT between two spheroidal dipoles of variable orientation, it is concluded that the generalized many-body approach for NFRHT developed here accurately accounts for geometric effects. As such, the generalized many-body approach can be applied to study new systems of nonspherical particles in the dipole limit.

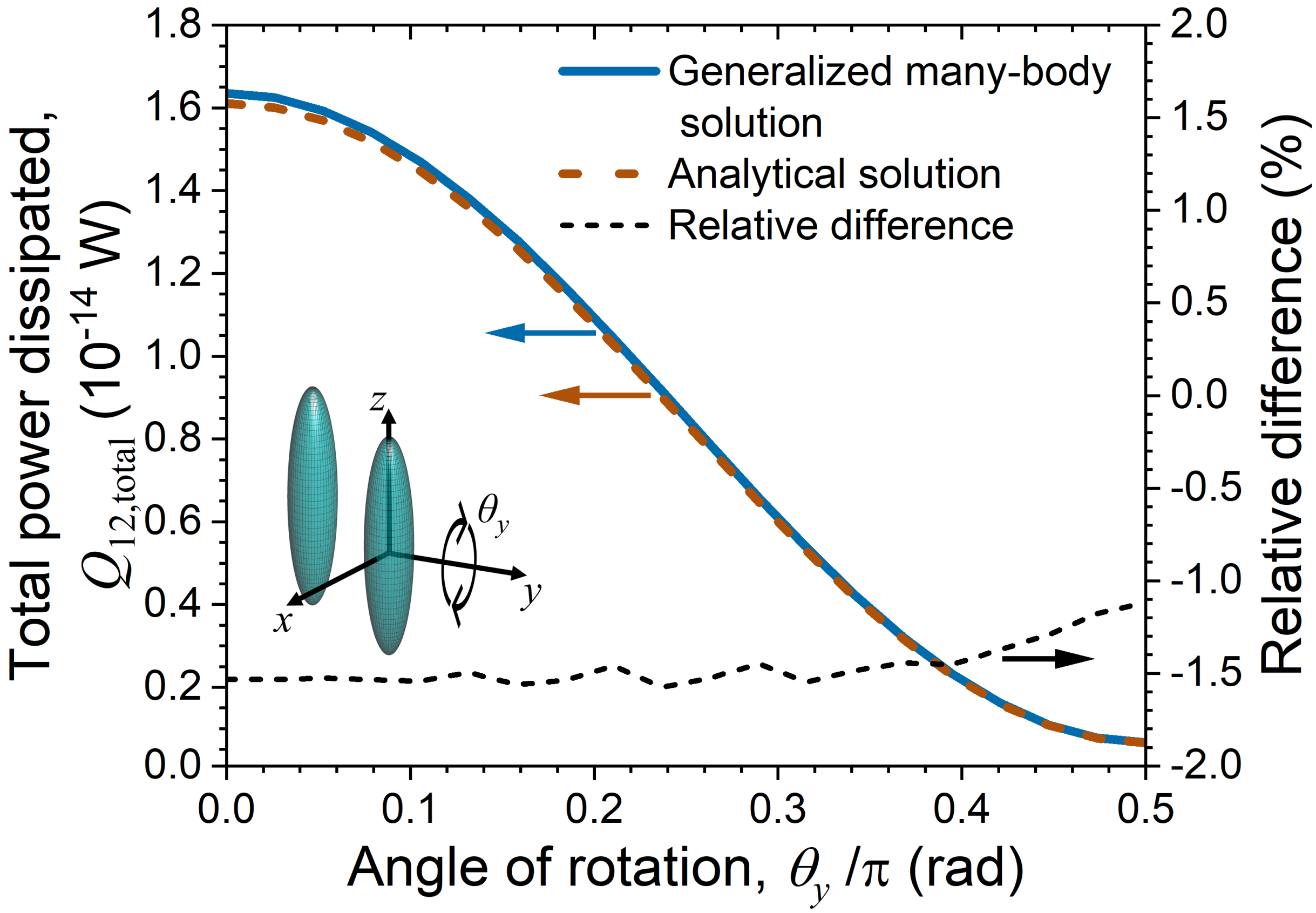

FIG. 4. Comparison of the generalized many-body approach against the analytical solution for total power dissipated in two SiC spheroidal dipoles of variable orientation. The dipoles are of dimensions $a = b = 15$ nm and $c = 75$ nm, separated by a center-of-mass distance $d = 7L_{\mathrm{ch}} = 525$ nm (separation distance not to scale in inset), embedded in vacuum ($\varepsilon_{\mathrm{ref}} = 1$), and at temperatures $T_1 = 0$ K and $T_2 = 300$ K. One particle is rotated around its local $y$-axis (see inset). The analytical solution is calculated from Ref. [24]. The relative difference between the generalized many-body solution and the analytical solution is calculated as $\left(Q_{12,\mathrm{total}}^{\mathrm{many\text{-}body}} - Q_{12,\mathrm{total}}^{\mathrm{analytical}}\right)/Q_{12,\mathrm{total}}^{\mathrm{analytical}}$.

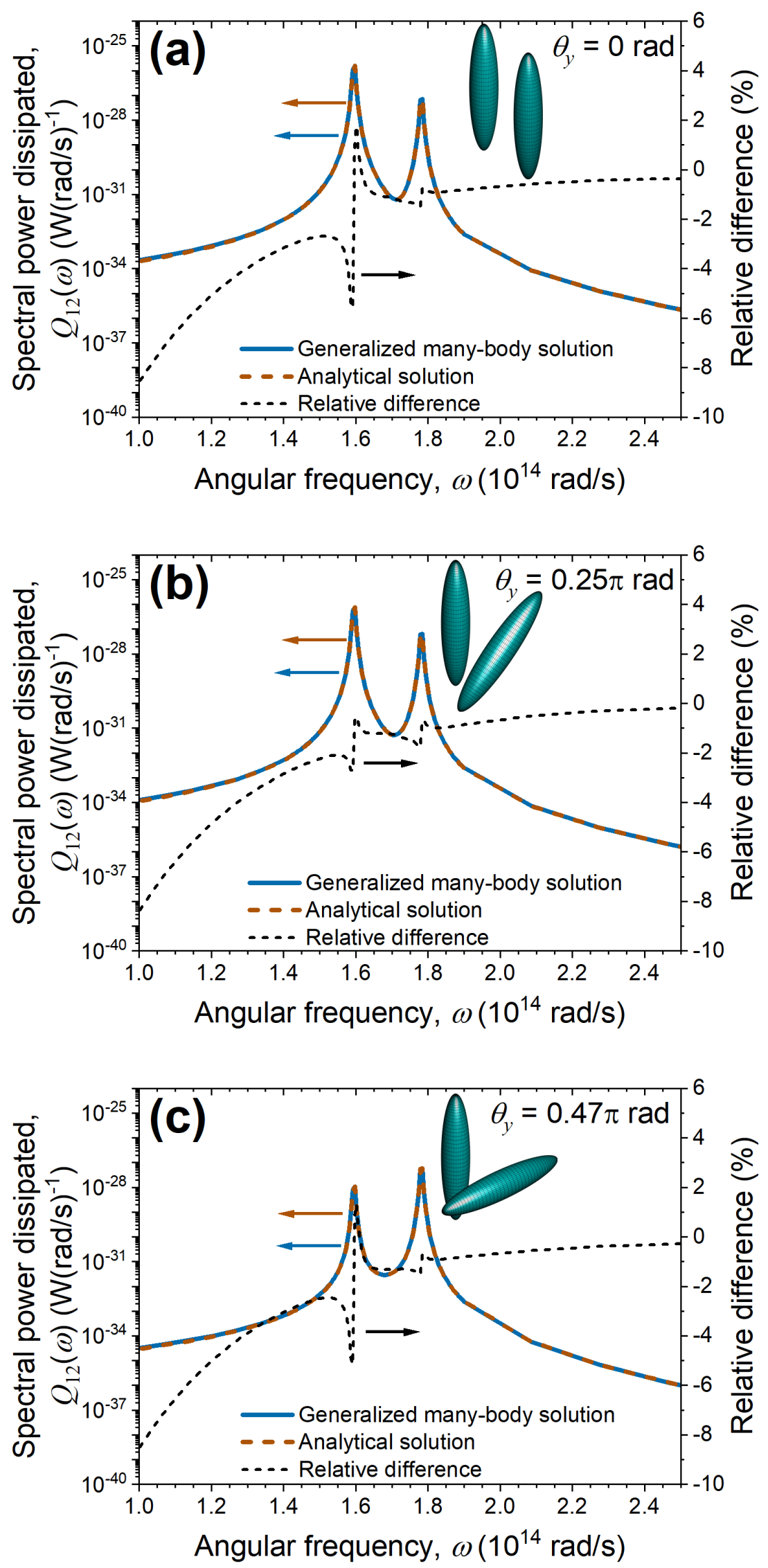

FIG. 5. Comparison of the generalized many-body approach against the analytical solution for spectral power dissipated in two SiC spheroidal dipoles. The dipoles are of dimensions $a = b =$ 15 nm and $c = 75$ nm, separated by a center-of-mass distance $d = 7L_{\text{ch}} = 525$ nm (separation distance not to scale in insets), embedded in vacuum ($\varepsilon_{\text{ref}} = 1$), and at temperatures $T_1 = 0$ K and $T_2 = 300$ K. The analytical solution is calculated from Ref. [24]. The relative difference between the generalized many-body solution and the analytical solution is calculated as $\left[Q_{12}^{\text{many-body}}(\omega) - Q_{12}^{\text{analytical}}(\omega)\right]/Q_{12}^{\text{analytical}}(\omega)$.

**B. Analysis of NFRHT between two ellipsoidal particles of variable configuration**

The generalized many-body approach is applied to study NFRHT in systems of two SiC ellipsoidal particles embedded in vacuum (Fig. 6). Both ellipsoidal particles are of dimensions $a = 15$ nm, $b = 45$ nm, and $c = 75$ nm. In the original configuration, one particle is translated a center-of-mass distance $d = 7L_{\mathrm{ch}} = 525$ nm along the $y$-axis from the position of the other particle located at the origin. Other configurations are created by rotating one particle in its local coordinate system by angles $\theta_x = \pi/2$, $\theta_y = \pi/2$, or $\theta_z = \pi/2$. In this way, four total configurations are analyzed, including the original unrotated setup.

NFRHT between particles is characterized by the spectral conductance calculated at a temperature of $T' = 300$ K. The spectral conductance between any two particles at temperature $T$ may be calculated as

$$G_{ij}(\omega, T) = \lim_{\delta T \to 0} \frac{Q_{ij}(\omega)}{\delta T} = \left[\frac{\partial \Theta(\omega, T')}{\partial T}\right]_{T'=T} \mathcal{T}_{ij}(\omega), \tag{37}$$

where $Q_{ij}(\omega)$ is the spectral power dissipated in particle $i$ due to radiative heat transfer with particle $j$,

$$Q_{ij}(\omega) = \left[\Theta(\omega, T_j) - \Theta(\omega, T_i)\right]\mathcal{T}_{ij}(\omega). \tag{38}$$

In the systems of ellipsoidal particles modeled, the spectral conductance displays splitting into three LSPh resonances [Fig. 6(d)–(f)]. These three resonances are correlated with the three unique semiaxis dimensions of the ellipsoids ($a \neq b \neq c$). Such spectral behavior follows that seen for SiC spheroidal particles [i.e., two LSPh resonances corresponding to two unique semiaxis dimensions ($a = b \neq c$), see Fig. 5] and for SiC spherical particles [i.e., one LSPh resonance corresponding to one unique semiaxis dimension ($a = b = c$), see Supplemental

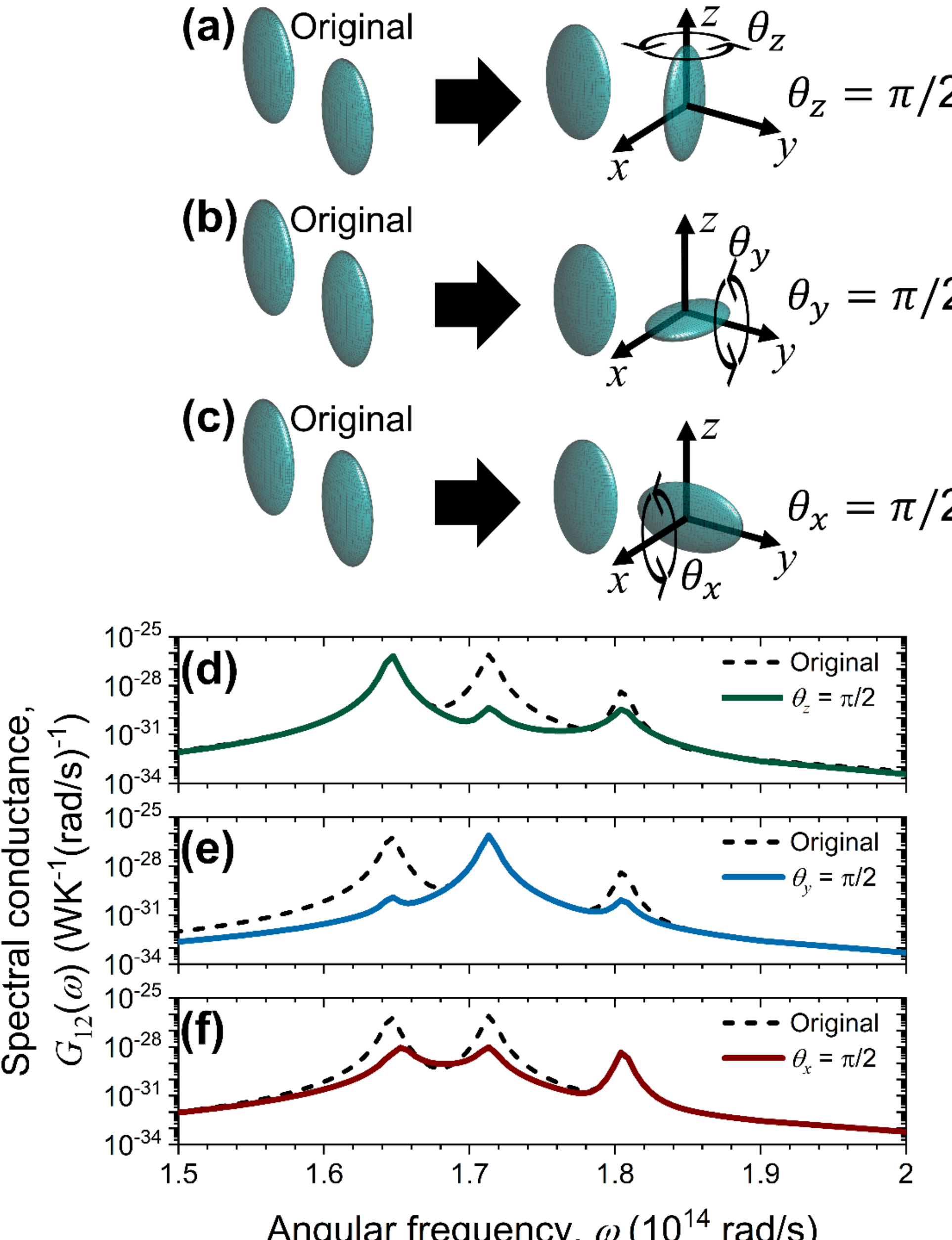

FIG. 6. Spectral conductance between two SiC ellipsoidal particles of variable orientation. One particle is rotated around its local coordinate system by an angle of (a),(d) $\theta_z = \pi/2$; (b),(e) $\theta_y = \pi/2$; or (c),(f) $\theta_x = \pi/2$ from the original system configuration. The particles are of dimensions $a = 15$ nm, $b = 45$ nm, and $c = 75$ nm, separated by a center-of-mass distance $d = 7L_{\text{ch}} = 525$ nm [separation distance not to scale in (a)-(c)], embedded in vacuum ($\varepsilon_{\text{ref}} = 1$), and the spectral conductance is calculated at temperature $T = 300$ K.

Material, Sec. S1, Fig. S1 [47]]. The frequencies at which LSPh resonances occur ($\omega_{\text{LSPh}}$) for an isolated ellipsoidal dipole can be estimated by determining the condition for which the weak form of the polarizability [Eq. (32)] diverges [41]:

$$\text{Re}\{\varepsilon(\omega_{\text{LSPh}}\}) = \varepsilon_{\text{ref}}\left(1 - \frac{1}{L_{i,\gamma}}\right), \tag{39}$$

where $\gamma$ represents Cartesian coordinates and $L_{i,\gamma}$ is given by Eqs. (25)–(27). For ellipsoidal dipoles of dimension $a$ = 15 nm, $b$ = 45 nm, and $c$ = 75 nm, the frequencies at which LSPh resonances are predicted to occur are, respectively, $1.647 \times 10^{14}$ rad/s, $1.713 \times 10^{14}$ rad/s, and $1.806 \times 10^{14}$ rad/s. These values match with excellent agreement to the resonance frequencies observed in the spectral conductance in Fig. 6(d)–(f), since the dipole size parameter is smaller than ~0.24, a regime in which the weak form of the polarizability provides accurate results (see Sec. III.C).

Rotation of one of the ellipsoidal particles results in thermal switching behavior in the spectral conductance. For configurations in which one particle is rotated by $\pi/2$ rad around one of its $x$-, $y$-, or $z$- axes, two of the LSPh resonances become damped while the resonance associated with the semiaxis along the axis of rotation remains unchanged. For example, the spectral conductance values at $1.713 \times 10^{14}$ rad/s and $1.804 \times 10^{14}$ rad/s for a particle rotated by $\theta_z = \pi/2$ rad are, respectively, three orders and one order of magnitude less than that of the unrotated particle, while the resonance at $1.647 \times 10^{14}$ rad/s remains unchanged. In this way, quasi-monochromatic heat transfer can be obtained at a desired LSPh frequency simply by changing the orientation of a particle. Additionally, the total, spectrally integrated conductance for rotated configurations is about half that of unrotated configurations [e.g., $G_{12} = 1.76 \times 10^{-15}$ WK$^{-1}$ for the original configuration and $G_{12} = 8.35 \times 10^{-16}$ WK$^{-1}$ for the $\theta_z = \pi/2$ rad configuration in Fig. 6(d)]. Such effects could be utilized, for instance, for dynamic thermal management of microelectronic devices. This thermal switching behavior can be attributed to system-level geometric effects on LSPh modes.

### C. Comparison of NFRHT between two ellipsoidal SiC particles using strong- and weak-form calculations

Next, strong and weak calculations of NFRHT are compared for systems of two ellipsoidal SiC particles. Eq. (18) is used in strong-form calculations and Eq. (23) is used in weak-form calculations. In strong-form calculations, a radius of $R_{d,i} = \min\{a, b, c\}/2$ is used for the finite spherical exclusion volume $V_{d,i}$.

A total of three geometrically similar two-dipole systems with variable size parameter ($X = 0.047$, 0.24, 0.47) are modeled. In these systems, one dipole is at temperature $T_1 = 0$ K and the other is at $T_2 = 300$ K. Dipole pairs are oriented as in the "Original" systems depicted in Figs. 6(a)–(c). Both the total $Q_{12,\mathrm{total}}$ and spectral $Q_{12}(\omega)$ power dissipated in dipole 1 due to interaction with dipole 2 are calculated [see Table 1 and Figs. 7(a)–(c), respectively].

TABLE I. Comparison of the strong and weak forms of the total power dissipated in a dipole for pairs of ellipsoidal SiC dipoles of various sizes

| **Size parameter of ellipsoidal dipoles $\boldsymbol{X}$** | **Ellipsoid dimensions $\boldsymbol{a}$, $\boldsymbol{b}$, $\boldsymbol{c}$ and center-of-mass separation distance $\boldsymbol{d}$** | **Strong form total power dissipated $\boldsymbol{Q_{12,\text{total}}^{\text{strong}}}$ (W)** | **Weak form total power dissipated $\boldsymbol{Q_{12,\text{total}}^{\text{weak}}}$ (W)** | **Relative difference in total power dissipated $\boldsymbol{\frac{\left(Q_{12,\text{total}}^{\text{strong}}-Q_{12,\text{total}}^{\text{weak}}\right)}{Q_{12,\text{total}}^{\text{strong}}}}$** |
|---|---|---|---|---|
| 0.047 | $a = 15$ nm<br>$b = 45$ nm<br>$c = 75$ nm<br>$d = 7L_{\text{ch}} = 525$ nm | $1.213 \times 10^{-13}$ | $1.248 \times 10^{-13}$ | -2.9% |
| 0.24 | $a = 75$ nm<br>$b = 225$ nm<br>$c = 375$ nm<br>$d = 7L_{\text{ch}} = 2625$ nm | $3.721 \times 10^{-13}$ | $4.045 \times 10^{-13}$ | -8.7% |
| 0.47 | $a = 150$ nm<br>$b = 450$ nm<br>$c = 750$ nm<br>$d = 7L_{\text{ch}} = 5250$ nm | $2.114 \times 10^{-12}$ | $4.500 \times 10^{-12}$ | -112.8% |

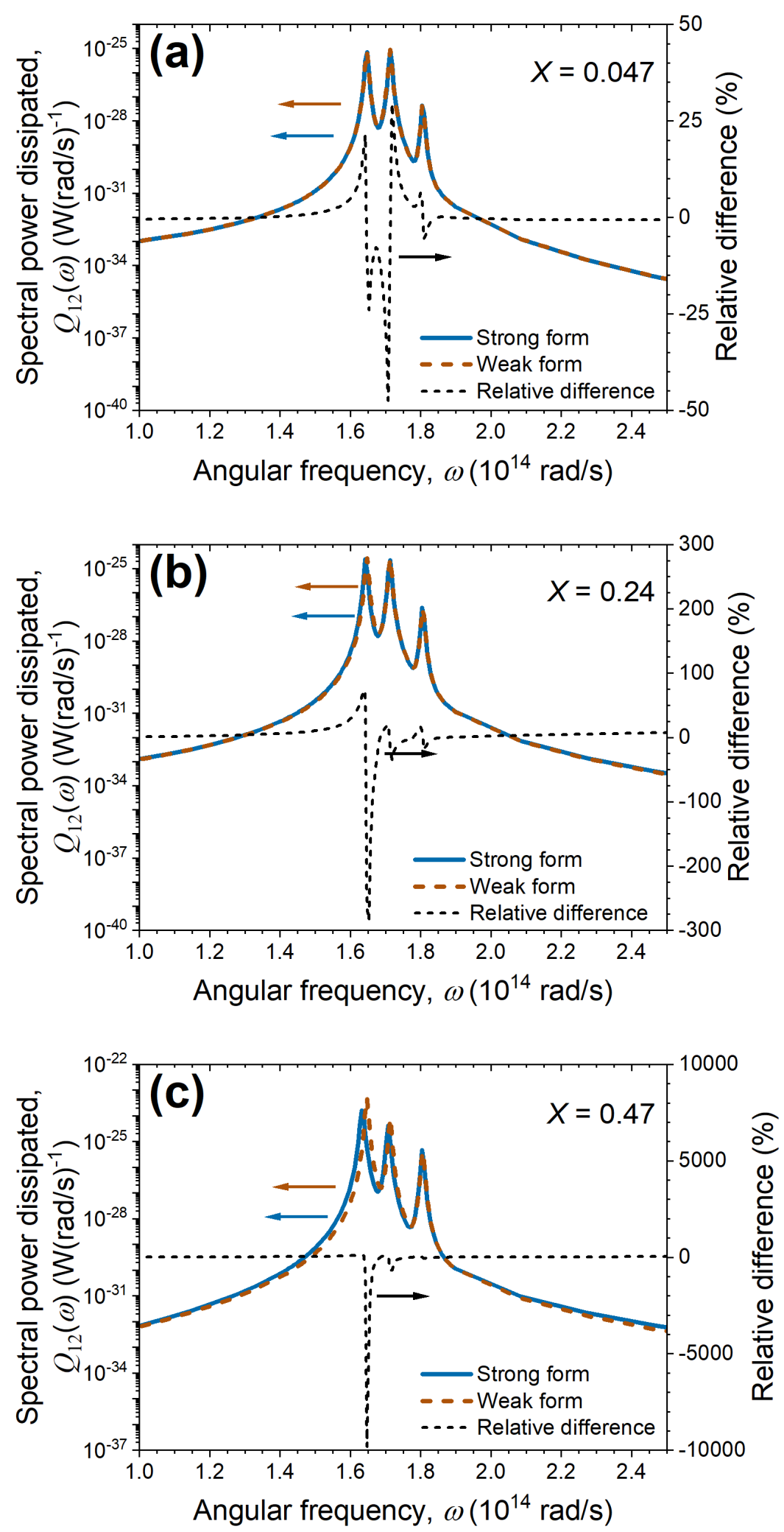

FIG. 7. Comparison of the strong and weak forms for the spectral power dissipated in pairs of ellipsoidal dipoles, with one dipole at temperature $T_1 = 0$ K and the other at $T_2 = 300$ K. Three geometrically similar systems of ellipsoidal dipoles are modeled: (a) dimensions $a = 15$ nm, $b = 45$ nm, $c = 75$ nm, size parameter $X = 0.047$, and center-of-mass separation distance $d = 7L_{\text{ch}} = 525$ nm; (b) dimensions $a = 75$ nm, $b = 225$ nm, $c = 375$ nm, size parameter $X = 0.24$, and center-of-mass separation distance $d = 7L_{\text{ch}} = 2625$ nm; and (c) dimensions $a = 150$ nm, $b = 450$ nm, $c = 750$ nm, size parameter $X = 0.47$, and center-of-mass separation distance $d = 7L_{\text{ch}} = 5250$ nm. The ellipsoidal dipoles are made of SiC and are embedded in vacuum ($\varepsilon_{\text{ref}} = 1$).

As predicted from the analysis of strong and weak forms of polarizability in Sec. III.C, the strong and weak forms agree well for systems in which the size parameter of constituent particles is less than ~0.24. Above this value, NFRHT calculations vary significantly between the strong and weak forms, following the same blueshift trends as seen in the polarizability analysis in Sec. III.C. In systems with size parameters greater than ~0.24, the strong form of NFRHT should be used.

### D. Analysis of the radiative thermal conductivity of a metamaterial composed of ellipsoidal particles

Lastly, the generalized many-body approach is applied to calculate the tensor spectral radiative thermal conductivity in a metamaterial composed of aligned $SiO_2$ ellipsoidal particles (Fig. 8). All particles are of dimensions $a = 15$ nm, $b = 45$ nm, and $c = 75$ nm and are arranged on a cubic lattice with lattice constant $d = 7L_{\mathrm{ch}} = 525$ nm. The spectral radiative thermal conductivity along the $x$-direction is calculated at temperature $T = 300$ K as [18]

$$\kappa_{xx}(T) = \frac{1}{A_{yz}} \sum_{i \in V_A} \sum_{j \in V_B} G_{ij}(T) \left( \mathbf{r}_j - \mathbf{r}_i \right) \cdot \hat{\mathbf{x}}. \tag{40}$$

Here, $V_A$ represents the group of particles on one side of a central dividing $yz$-plane, and $V_B$ represents the group of particles on the other side of the plane. More information on this calculation can be found in Ref. [18]. The cross-sectional area $A_{yz}$ of the central dividing $yz$-plane is calculated as $A_{yz} = N_y N_z d^2$, where $N_y$ and $N_z$ are the number of particles along the $y$- and $z$-directions, respectively. Calculation of the radiative thermal conductivity along the $y$- and $z$-directions is found through cyclic permutation of Cartesian coordinates in Eq. (40). Based on the convergence analyses for spherical $SiO_2$ particles in Ref. [18], we simulate metamaterials with $N_x = N_y = N_z = 10$ particles, such that the model comprises $N_x \times N_y \times N_z = 1000$ particles. The dielectric function of $SiO_2$ is calculated using a multi-oscillator Lorentz model,

$$\varepsilon(\omega) = \varepsilon_{\infty} + \sum_{n=1}^{3} \left[ \frac{S_n}{1 - \left(\frac{\omega}{\omega_{0,n}}\right)^2 - i\Gamma_n \left(\frac{\omega}{\omega_{0,n}}\right)} \right], \tag{41}$$

with parameters given by Ref. [48].

At the low LSPh resonance around $9 \times 10^{13}$ rad/s, the spectral radiative thermal conductivity at resonance in the $y$- and $z$-directions are, respectively, 2.0 and 2.8 times that in the $x$-direction. At the high LSPh resonance around $2.1 \times 10^{14}$ rad/s, the spectral radiative thermal conductivity at resonance in the $y$- and $z$-directions are, respectively, 1.6 and 2.0 times that in the $x$-direction. In this way, anisotropic radiative thermal conductivity may be achieved and controlled through choice of shape of constituent particles in a metamaterial. In Fig. 8, there are also slight shifts in the frequency of resonance for radiative thermal conductivities of different directions compared with the resonance frequencies for a metamaterial composed of spherical particles.

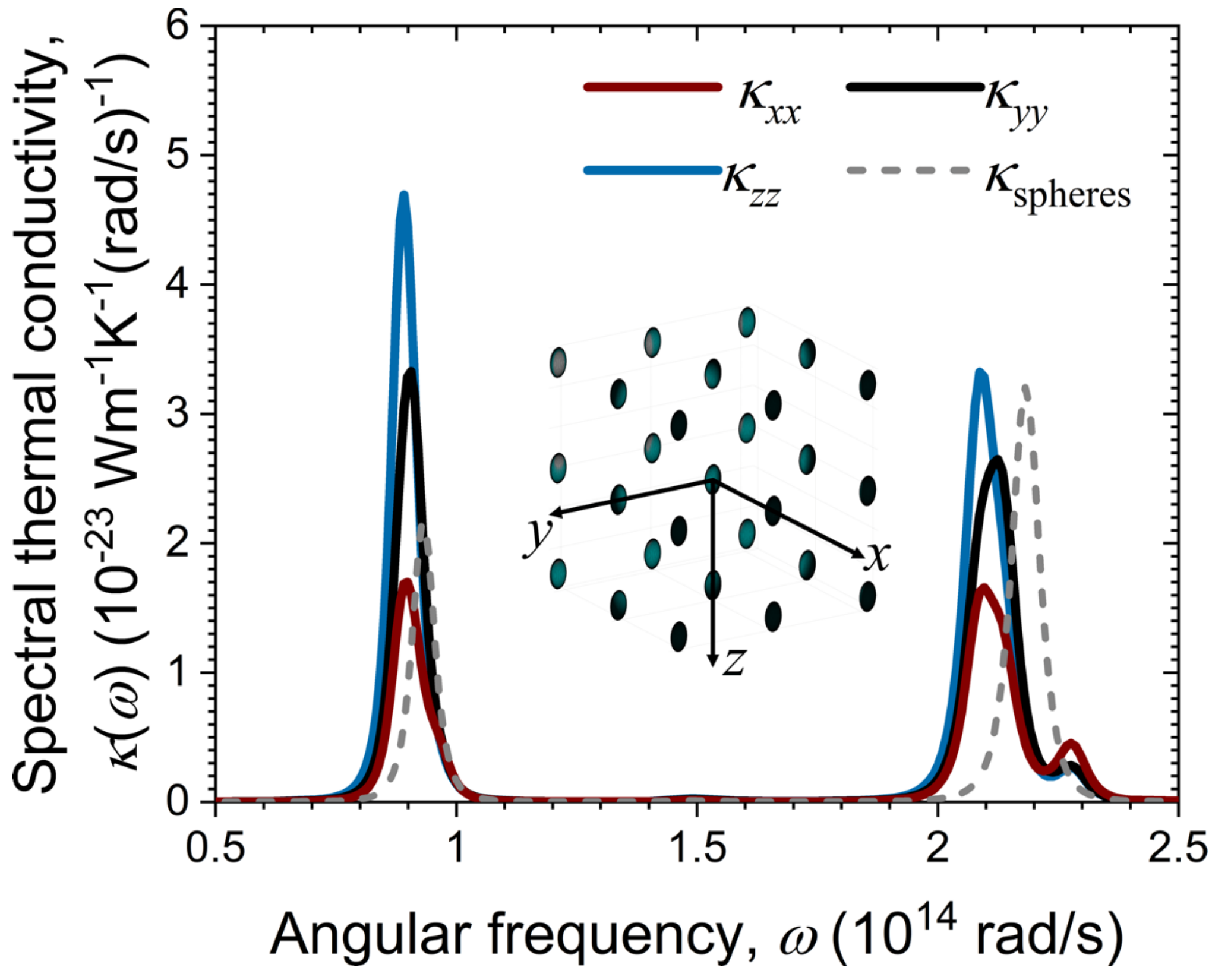

FIG. 8. Spectral radiative thermal conductivity of a metamaterial constructed of ellipsoidal $SiO_2$ particles embedded in vacuum ($\varepsilon_{\mathrm{ref}} = 1$). Particles are of dimensions $a = 15$ nm, $b = 45$ nm, $c = 75$ nm and are arranged on a cubic lattice with lattice constant $d = 7L_{\mathrm{ch}} = 525$ nm. The inset schematic displays a 3×3×3 sample of the full 10×10×10 metamaterial represented by the data. The solution for a metamaterial constructed of spherical $SiO_2$ particles of equivalent volume and equivalent lattice constant is shown by the dashed line. The spectral radiative thermal conductivity is calculated at temperature $T = 300$ K.

## V. CONCLUSIONS

In this paper, we developed a generalized many-body approach for calculating NFRHT between nonspherical dipoles. Based upon the self-consistent system Green's function formulation common to many-body approaches, the unique geometric parameters of nonspherical dipoles were implemented in the definition of the self-term of the free-space Green's function. Both strong and weak forms of the generalized many-body approach for NFRHT were presented for particles in the dipole limit. The strong form resulted in the most accurate models of NFRHT across particle length scales, whereas the weak form was less computationally expensive but only applicable to very small particles (i.e., particles with size parameter less than ~0.24).

In the generalized many-body approach presented here, dipole polarizability is defined *a posteriori* from the free-space Green's function solution. Calculation of polarizability is an optional post-processing step rather than a required input. It is shown that the polarizability defined for nonspherical dipoles presented here may be implemented as-is within existing dipole models of NFRHT that use polarizability as an input parameter.

We compared the generalized many-body method against an analytical solution for NFRHT between two spheroidal dipoles. A good agreement was obtained. The small discrepancies between results of the two methods arise from differences in approximations for multiple reflections.

We then applied the generalized many-body method to analyze the near-field spectral conductance between two SiC ellipsoidal dipoles. It was found that ellipsoidal dipole geometries resulted in a splitting of the spectral conductance into three unique resonances correlated with the LSPh modes of ellipsoidal dipoles. This spectral behavior contrasts with the single LSPh resonance observed for NFRHT between spherical dipoles and is attributed to the unique semiaxis dimensions that define ellipsoidal dipole geometries. Changes in the orientation of one of the ellipsoidal dipoles resulted in independent, active tuning of LSPh resonance by up to three orders of magnitude. Such tuning capabilities could be utilized for engineering thermal management strategies in microelectronic devices.

Finally, we calculated the spectral radiative thermal conductivity of a metamaterial composed of $SiO_2$ ellipsoidal particles. We found that the metamaterial displayed anisotropic radiative thermal conductivity, with differences in the value at resonance up to 2.8 times between different directions.

For future work, it is recommended that the generalized many-body method be applied to study NFRHT between other nonspherical dipole shapes, such as cylinders and rectangular parallelepipeds. In addition, the many-body method is currently limited to electric point dipoles. For metallic nanoparticles, the magnetic dipole contribution is typically non-negligible. Magnetic dipole contribution could be added in the generalized many-body method by combining the procedure described in Sec II.C with the coupled electric and magnetic dipole approach developed by Dong et al. [49] for spherical nanoparticles.

## ACKNOWLEDGMENTS

This work was supported by the National Science Foundation (Grant No. CBET-1952210). L.P.W. acknowledges that this material is based upon work supported by the National Science Foundation Graduate Research Fellowship under Grant No. DGE-1747505. Any opinions, findings, and conclusions or recommendations expressed in this material are those of the authors and do not necessarily reflect the views of the National Science Foundation. The support and resources from the Center for High Performance Computing at the University of Utah are gratefully acknowledged. Additionally, the authors would like to thank Dr. Matthias

Krüger for clarifying Eq. (4) of Ref. [24], and Lívia Corrêa McCormack for supporting figure development in this manuscript.

# Supplemental Material

## Generalized many-body approach for near-field radiative heat transfer between nonspherical dipoles

Lindsay P. Walter and Mathieu Francoeur*

Department of Mechanical Engineering, The University of Utah, Salt Lake City, UT 84112, USA

*Corresponding author: mfrancoeur@mech.utah.edu

## S1. SPECTRAL CONDUCTANCE OF TWO SPHERES

The spectral conductance between two SiC spheres is calculated using the generalized many-body approach. Eq. (15) of the main manuscript is used to define the self-interaction free-space Green's function. Results are shown in Fig. S1.

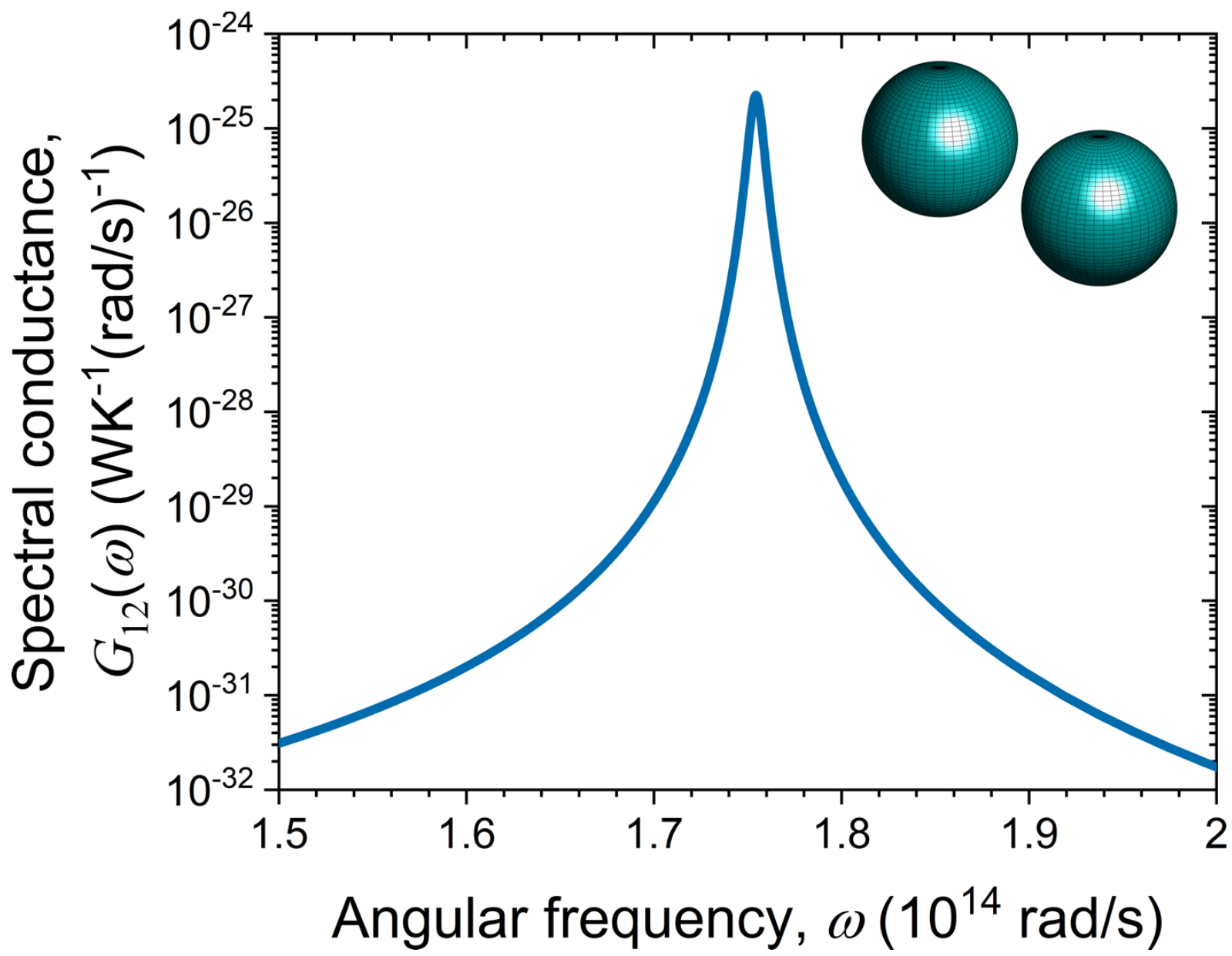

FIG S1. Spectral conductance between two SiC spherical particles. The particles are of dimensions $a = b = c = 35$ nm, separated by a center-of-mass distance $d = 7L_{\mathrm{ch}} = 245$ nm (separation distance not to scale in inset), embedded in vacuum ($\varepsilon_{\mathrm{ref}} = 1$), and the spectral conductance is calculated at temperature $T = 300$ K.